\documentclass[aps,prl,preprint,groupedaddress,showpacs, showkeys]{revtex4}

\begin{document}


\title{A Natural Value Unit - Econophysics as Arbiter between Finance and Economics}


\author{Steivan Defilla}
\email[]{steivan.defilla@seco.admin.ch}
\thanks{this research is a personal view and does not reflect the position of SECO. Acknowledgements: Hans Wolfgang Brachinger (University of Fribourg), Daniel T. Spreng (Center for Energy Policy and Economics, ETH Zurich), Swiss Federal Statistics Office and SECO, Bern.}
\affiliation{Energy Advisor, Swiss State Secretariat for Economic Affairs SECO}


\date{\today}

\begin{abstract}
Foreign exchange markets show that currency units (= accounting or nominal price units) are variables. Technical and economic progress evidences that the consumer baskets (= purchasing power units or real price units) are also variables. In contrast, all physical measurement units are constants and either defined in the SI (=metric) convention or based upon natural constants (= "natural" or Planck units). Econophysics can identify a constant natural value scale or value unit (natural numeraire) based upon Planck energy. In honor of the economist L. Walras, this "Planck value" could be called walras (Wal), thereby using the SI naming convention. One Wal can be shown to have a physiological and an economic interpretation in that it is equal to the annual minimal real cost of physiological life of a reference person at minimal activity. The price of one Wal in terms of any currency can be estimated by hedonic regression techniques used in inflation measurement (axiometry). This pilot research uses official disaggregated Swiss Producer and Consumer Price Index data and estimates the hedonic walras price (HWP), quoted in Swiss francs in 2003, and its inverse, the physical purchasing power (PhPP) of the Swiss franc in 2003.  
\end{abstract}


\maketitle

\section{1. The anecdote of the cheated cheater}

\emph{A baker once went to the judge to complain that the farmerÕs chunks of butter were smaller and smaller every year, which according to him meant that the farmer was cheating and should be fined. Questioned by the judge, the farmer replied that he has not noticed any change and that he was obliged to use the bakerÕs bread loafs as counterweights for setting the weight of his chunks of butter as he had no other counterweights, and that if anyone was cheating, it was the baker}.
\par 
This anecdote illustrates the circular measurement paradigm and its variable and imprecise measurement units that are symptomatic for finance and economics. Unlike all other sciences, which clearly differentiate between the measuring phenomenon (i.e. the fix measurement unit) and the measured phenomena (i.e. variable measurands), finance and economics may easily confuse the two, so that it is not always clear what is measuring and what is being measured. While currency units are normally fix \emph{nominal} price units, they turn out to be variables on foreign exchange markets. Likewise, consumption baskets are the fix \emph{real} price units in inflation analysis, but they are variables in consumption analysis. Without a fix and precise value measurement unit, value accounting is like adding apples to tomatoes and becomes meaningless. Without meaningful accounting, finance, economics and econophysics would not be quantitative sciences. From the metrological point of view it is not the diversity of measurement units, but their \emph{variability} that creates the problem:  A length unit such as the \textquotedblleft King's arm length\textquotedblright can perfectly coexist with the meter, but it might be too imprecise for science and trade if, e.g., it gains a certain percentage of length every year, as the King grows. Contrary to relativity theory, where the varying apparent length of the meter in movement is the result of a precisely defined and well-measured physical phenomenon, the variability of measurement units in finance and economics is the result of a poorly defined measurement paradigm.
\par
Finance and accounting still ignore the variability and imprecision of the money unit to a large extent. Financial statements and accounts are held in nominal currencies, income taxes are paid on nominal income, VAT on nominal prices, dividends calculated on the basis of nominal profits, fines stated as nominal amounts. For finance, money counts more than anything else, money is the numeraire. In terms of the above anecdote, the accountant is the farmer, as he does not notice any drift in time. For the accountant, the optimal strategy to get rich would be to print money. The hyperinflation in Germany after World War I has disproved that strategy, as it made Germany really poor, albeit astronomically high nominal GDP figures. Fisher\cite{Fisher1928} and Keynes\cite{Keynes1936}  called this view money illusion. Value theories, e.g. by Debreu\cite{Debreu1959}, normally ignore money illusion.

\section{2. Consumer basket illusion and index inconsistency}

In economics, the consumer basket is the numeraire. This allows calculating real exchange rates, also called purchasing power parities (PPP) between currencies, and inflation, i.e. loss of purchasing power in a given time span, where from real prices, real income, real profits, real GDP, real taxes and real fines can be computed. The consumer basket is not an ideal numeraire because it is not unique. In fact each individual has his own consumer basket. Economics understands the composition of each individual consumer basket as the result of constrained optimization, e.g. cost minimization under production constraint or utility maximization under budget constraint, cf. e.g. pp. 400 ss. of \cite{Chiang1984}. These optimization results depend on relative prices and are therefore \emph{measurement results}, corresponding to subjective valuation of goods and services, or to subjective purchasing power of money, but not an objective value \emph{measurement unit}.
\par
For getting rid of individual basket variability, economics uses the \emph{aggregate basket} as numeraire. Even the aggregate world basket varies however in its composition in time as a function of technological innovation, relative prices, global economic development level, global average income and global income distribution, global ethnic composition and global climate change. Furthermore, individual items undergo quality change in time\cite{Pollak1983}. Only the sum of all world baskets over all the years is constant, but it is known only at the End of History. This shows that the aggregate basket approach is either not practicable or, if it is, it confuses a value measurement unit with a measurement result of \emph{collective} or \emph{aggregate} valuation of goods and services, i.e. with aggregate purchasing power of money, which is an average. Any average is highly dependent on the data of its constitutive set. This confusion can be called \emph{consumer basket illusion}. In practice, no national statistics authority has ever calculated national inflation on the basis of the world basket. The choice of the basket remains to some extent a political one. Neither does the purchasing power parity (PPP) calculation of the World Bank inspire sufficient confidence to the World Bank for using it to calculate its Members\textquoteright \space quotas.
\par
This lack of confidence may also be related to the fact that the ideal multilateral index method does not exist. The bilateral Laspeyres index\cite{Laspeyres1871} e.g. has been shown to considerably \emph{overstate} inflation and \emph{understate} deflation (i.e. this method is not reversible, it is systematically bullish). The Boskin Commission Report\cite{Boskin1996} has concluded that the Laspeyres index overestimated the annual US inflation during the past decades by 1.1 percent per year due to four sources of upward bias: Substitution bias, quality bias, new product bias and outlet bias. Similarly, the bilateral Paasche index\cite{Paasche1875} is systematically bearish. The bilateral Fisher index\cite{Fisher1922} does not have systematic bias but fails nonetheless to be transitive, i.e. to create a unique consistent cardinal ordering of countries in a multilateral context. This means it is sensitive to the choice of a base country and base year. 
\par
The above problems arise if the indices are applied as \emph{star indices}. A star index has a fix base year or base country. For inflation, \emph{chain indices} have become increasingly popular. They have no base year and link a year only with its immediate neighbors. Thus they yield a seemingly unique year-on-year inflation, whose cumulation over many years gives however a different inflation from the corresponding star index over the same period.  For international comparisons, chain indices are not appropriate, because countries are not naturally ordered in a chain. A great number of multilateral indices have therefore been proposed\cite{Summers1991} \cite{Neary2004}; some are transitive, if the set of countries or years is fix. The literature on axiomatic and economic approaches to index number calculations\cite{Diewert1996} \cite{Hill2004} shows however that without any restricting hypothesis, all indices have an inconsistency problem: Either the ordering they create is sensitive to the choice of the base country or the base year, or it is sensitive to the inclusion of further countries or years, or it is sensitive to the inclusion of further goods and services, or it is sensitive to the substitution behavior (i.e. the specification of the utility or production function) the indices implicitly imply. 
\par
In terms of the above anecdote, the economist is the baker. He has the advantage over the farmer that his bread conserves better in time than the farmerÕs butter. Possibly he uses the farmerÕs butter chunks as weight standards for his bread size. But he produces many different bread sizes, depending on demand. Each month he keeps one loaf of each size as reference. This leaves him with too many weight units, each giving different measurement results. Since Walras\cite{Walras1874}, economists believe that any good or service can be chosen as numeraire. Less understandable are some economists\textquoteright \space opinions that defining a value (or real price) unit is useless or impossible. No value theory allows baskets to be chosen as numeraires, yet this corresponds to established practice in inflation measurement and in international purchasing power comparisons.

\section{3. Optimal numeraire choice}

\par
\newfont{\mengen}{dsrom10}
Econophysics is ideally placed to play the role of the judge in the above anecdote by defining a fix and precise measurement unit in order to determine the conversion factors of the existing units. Due to the abstract nature of value, the conceptual challenge is however greater than for usual physical problems. The 18$^{th}$ century Rationalists (Spinoza, Leibnitz) doubted that value was at all objective and measurable (\textquotedblleft value judgement\textquotedblright  \space as opposed to measurable and objective fact). A measurement in general is any process associating a given phenomenon with a unique real number. The most important measurement process of economics is the market. The minimal requirement for value is to be a real number ($\in \mengen R$), positive for goods, negative for \textquotedblleft bads\textquotedblright.  Value is however not an intrinsic or absolute property of a good, but an \emph{extrinsic} or \emph{interactive} property. Value is always the result of an \emph{interaction} between a specific good and a specific agent: A gun is a good for the owner and a bad for the victim. If their valuations disagree, it is because they have different interactions with the gun. An example of an interactive physical property of an object is its weight, as it depends on the surrounding gravitation. The value or price of a good is highly interactive, as it depends not only on the intrinsic qualities of the good, but also on the buyer and his tastes and preferences, on the general level of purchasing power of buyers, on the price of competing goods, as well as on the interaction of the good with other goods, i.e. on the physical and chemical environment of the good. A fridge is of less value to the Eskimos than to the Kuwaitis. Water may be a good of first necessity, but if a house is flooded, water turns into a bad due to its uncontrolled interaction with other goods. Marx erred in his belief that value was intrinsic to goods (in centrally planned economies such as in the former Soviet Union, the price of the goods used to be engraved in the good). Not even money has an absolute value, in spite of a face value being printed on it. The transaction value of money is called purchasing power and is an interactive quantity that depends on the supply and velocity of money and on what goods and services the money holder effectively buys (quantity theory of money)\cite{Fisher1911}. 
\par
A consequence of the interactive character of value is the fact that the marginal value (or marginal utility) of goods is normally decreasing ($\frac{\partial V}{\partial Q}<0$). The $101^{st}$ loaf of bread consumed in a month has less value than the $100^{th}$, p. 14 of\cite{Fisher1892}. The decreasing marginal utility of goods is the result of consumers\textquoteright \space limited consumption capacity, which means that consumers prefer consuming goods in an appropriate - but always changing - mix. A numeraire (value unit or real price unit) by definition must have a constant marginal value ($\frac{\partial V}{\partial Q}=0$), otherwise no accounting would be possible. In the absence of meaningful accounting, finance, economics and econophysics would not be quantitative sciences. Economics \emph{by hypothesis} considers the marginal value of consumer baskets to be constant. In reality, it is decreasing, too, which is evidenced by the fact that an individual's consumer basket changes composition with increasing wealth. 
\par
The success story of physical methods lies in the fixity and precision of its measurement units. The better the units, the better the instruments, and the better the results. In physics, the same measurement units are valid for all theories and models. If econophysics is to abide by this almost axiomatic principle, it must start identifying a numeraire whose constant marginal value is verified \emph{by virtue of an interactive physical phenomenon}. This means identifying a type of interaction that can become a  \emph{fix}  and \emph{precise} value unit by means of which all valuations can be expressed. Contrary to the Gold Standard, which was a man-made law, the physical numeraire exists independently of man-made laws. In monetized economies it should at any time and place be possible to calculate the exchange rate or purchasing power of the used currency against that numeraire, an exchange rate that is variable in time and place. Given the genuinely physical character of this kind of numeraire, it is convenient to call the purchasing power of money in terms of that numeraire the physical purchasing power (PhPP), in order to differentiate it from the usual social purchasing power of money. 
\par
As measurement units must always be invariant with respect to the phenomena they measure, it now remains for econophysics to identify which interaction could be a value unit (i.e. a numeraire) that is invariant to the individual or aggregate valuation of economic agents. An ideal numeraire choice has then to satisfy the following criteria: 

\par 1) A given quantity of numeraire interacting with a given environment must have a unique value based on an interactive physical phenomenon. In particular, this value must be different from the numeraire\textquoteright s production cost and its market price and also be different from any subjective, collective or aggregate valuation of that quantity of numeraire, i.e. it must be \emph{physical} value. Only the numeraire must be bought for its physical value.

\par 2) The numeraire must be a scalar. Weighted consumer baskets or vector type numeraires are always measurement results. Without restrictive hypotheses they do not produce unique consistent multilateral orderings of the countries\textquoteright \space real incomes, as the impossibility to find ideal multilateral indices shows.

\par 3) The numeraire must not have quality variations that influence its value. This excludes the choice of the labour hour as numeraire, as it has extreme quality variations:  In one hour, one can either win several hundreds of millions of dollars in a lotto, or else one can destroy the whole planet by firing weapons of mass destruction. The value of a labor hour is one of the most interesting \emph {variable} phenomena of economics.

\par 4) The numeraire must be available on a market and must not have a zero market price. If it had a zero market price, it could not be used for measuring the purchasing power of any currency, as this would entail dividing nominal prices by zero. This criterion excludes taking non-marketed goods such as  air, sea water and subsoil natural resources as numeraire. 

\par 5) The numeraire must be a one-time consumable. Only one-time consumables may have a unique value when interacting with a reference environment. The value of durables (e.g. real estate) is not unique (it depends on how the agents use them) and not known before the end of their life cycle (e.g. for real estate in the infinitely far future). This criterion also excludes the choice of gold as numeraire. 

\par 6) The numeraire must not be a purely man-made product, which might be altered or disappear if its production is stopped. Such a numeraire would have a limited temporal and geographic validity. If it is too specific and produced by a single enterprise, its price might be set by a small group of persons, hence its use as numeraire might be subject to manipulation. This is the main problem of choosing the Big Mac numeraire, cf. $http://www.economist.com/markets/Bigmac/Index.cfm$, which would satisfy all the other criteria. 

\par 7) The numeraire must be part of the consumption basket of all economic agents independently of their wealth level, and of all human societies independently of their economic development level. Value is of physio-biological nature, as it results from the interaction of biological organisms with their environment in order to acquire their resources.  

\par
In physical science there exists one concept that satisfies all the above criteria. It comes from engineering and has originally been called  \emph{available energy} by Gibbs\cite{Gibbs1873}. More recently it has been generalized and called  \emph{exergy} by Z. Rant\cite{Rant1956} in order to clearly differentiate it from internal energy. Internal energy is an intrinsic state function, whereas available energy or exergy is an extrinsic or interactive state function. Exergy or available energy is the maximum useful work a system can theoretically yield in a \emph{reversible} process when interacting with a precisely defined environment that is in physical and chemical equilibrium and  \emph{actively contributes} to the process, cf. pp. 150 ff. of\cite{Baehr2005}. A closed system in physical disequilibrium with respect to its environment has an exergy corresponding to: $E =W_{rev}= \Delta U+ p_a \Delta V -T_a\Delta S >0$, where the $\Delta U$, $\Delta V$ and $\Delta S$ denote respectively the difference in internal energy U, volume V and entropy S of the system between its disequilibrium state (p, T) and the environmental equilibrium state $(p_a,T_a)$. In a stationary working system, the instantaneous exergetic power $- P_{rev}$ lost by the system is given by $\dot {E}= - P_{rev} = \dot {Q} (T-T_a)/T$ where $\dot {Q}$ is the heat flow. A district heating system might be an example of such a system. This system produces heat of temperature T. The theoretically maximal work of its output is limited by the Carnot factor ${(T-T_a)/T}$.  In an open stationary system exchanging matter at a rate $\dot {m}$ with its environment, the instantaneous exergetic power lost by the system generalizes to $\dot {E}= - P_{rev} = \dot {Q} (T-T_a)/T  +\dot {m} (h-h_a - T_a(s-s_a))$, where h is the specific enthalpy and s the specific entropy of the flowing substance.
\par
Exergy or available energy is also present in chemical reactions. The maximum work of chemical reactions is given by the Gibbs function of the corresponding reversible isobar-isotherm reaction, which is negative for a combustion: $E = -W_{rev}= -  \Delta ^R G(p_a, T_a)<0$. Reversible isobar-isotherm reactions are approximated in \textquotedblleft cold combustion\textquotedblright (e.g. fuel cells, accumulators, energetic reactions taking place within living organisms). We have: $-  \Delta ^R G(p_a, T_a)= -\Delta ^R H(p_a, T_a) + T_a\Delta ^R S(p_a, T_a)= -\Delta ^RU(V) -p_a \Delta ^R V + T_a\Delta ^R S(p_a, T_a)$, where $-\Delta ^R H(p_a, T_a)$ is the isobar gross calorific value $GCV(p_a)$ of the combustion, $T_a\Delta ^R S(p_a, T_a)$ is the entropy production of the combustion at equilibrium conditions multiplied by the equilibrium temperature, $-p_a \Delta ^R V$ the volume variation at equilibrium pressure and $-\Delta ^RU(V)$ the isochoric gross calorific value $GCV(V)$ of the combustion measured experimentally by combustion in the calorimetric bomb in an artificial atmosphere of pure oxygen at high pressure (40 bar). This procedure is defined in several international standards: ISO standard 1928:1995 for solid fuels, the DIN 51900-1:2000-04 standard for liquid fuels. 
\par
The Gibbs function gives the reaction-specific exergy or available energy independently of its environment. It supposes that the reactants enter the reaction chamber separately, that the products leave the reaction chamber unmixed and that the reaction performs completely. Exergy or available energy exists also in reversible mixing with the physical and chemical equilibrium environment at $p_a, T_a$. After mixing the exergetic value (EXV) of products is equal to zero. Theoretically, the exergy of mixing could be used in osmotic power stations in which e.g. sweet water of a river is kept separated from the salt water of the sea by a semipermeable membrane. This kind of osmotic work is the proof that scarcity value is a physical concept. The chemical environment determines when the reaction reaches equilibrium. The exergetic value (EXV) of reactants can be calculated from the Gibbs function with a correction term: $E_x(p_a, T_a)= -  \Delta ^R G(p_a, T_a) + \Delta E(p_a, T_a)$. The correction term $\Delta E(p_a, T_a)$ depends on the relative abundance of the reactants and products when they interact with the equilibrium environment. This is illustrated in the following paragraph.  
\par
Ahrends\cite{Ahrends1977}\cite{Ahrends1980} found that in exergetic analysis the standard environment must be defined not only for its p and T, but also for its chemical and physical composition, and that it must be in equilibrium. Several attempts have been made to describe an earth-like equilibrium\cite{Ahrends1980}. Diederichsen\cite{Diederichsen1991} determines the equilibrium by minimizing the Gibbs potential over all the phases of the 17 most common elements and 971 most common compounds of the terrestrial atmosphere, hydrosphere and lithosphere. He found an earth-like equilibrium if the following three parameters were fixed: The temperature $T_a= 298.15 K$, the lithosphere depth $z_l = 0.1 $m and the hydrosphere depth $z_h=100$m, beyond which these two spheres are ignored. The equilibrium, i.e the point of overall minimal Gibbs potential, resulted at an equilibrium pressure $p_a = 91.771$  kPa, being the pressure of the entire atmosphere on all the condensed phases. This pressure that is only slightly lower than the 100 kPa standard pressure. Also the proportion of oxygen was slightly lower in the standard equilibrium atmosphere (16.5 percent) than in reality (23 percent). In either case, the EXV of oxygen amounted at most to one or two percent of the isobar GCV of a combustible hydrocarbon. If however the lithosphere depth parameter of $z_l = 0.1 $ m had been chosen to be $z_l= 1 m$ instead, the lithosphere would have consumed all the free oxygen. Oxygen would be very rare and its exergetic value EXV would then be a multiple of the EXV of hydrocarbons.  
\par
To sum this up, exergy or available energy can be calculated in practically any kind of process, provided simply that the process is specified as reversible (i.e. ideal) process. Reversible exergy-conserving processes represent the invariant scale against which process losses can be measured. These losses can be measured because exergy or available energy is a one-time consumable, contrary to total energy that is conserved in time and to entropy that is increasing in time in a closed system. In the basic definition of exergy given above, entropy is taken with negative sign as negentropy. The term ${T_aS}$ is called anergy, thus anergy + exergy = energy. If ${T_a}$ is fix, anergy is increasing in time, proportionally to entropy. The exergy concept combines the first and the second thermodynamic principles. This differentiates an exergetic numeraire from a negentropic numeraire that would only use the second thermodynamic principle. Geogescu-Roegen\cite{GeorgescuRoegen1971} concentrated on the second principle only. For a comprehensive bibliography on exergy see  $http://exergy.se/$. 
\par
It would be wrong to conclude from here that exergy is value. Even though both are interactive, value increases in the economic process whereas exergy decreases. Furthermore, the value of durable or investment goods is linked to their productivity, which is a concept totally different from exergy. A physical concept of productivity has been defined as catalytic activity. The corresponding SI unit is called katal, defined as mol/s. Value is more general than exergy. Exergy is not the numeraire either. If exergy was chosen as numeraire, all real wealth, which is normally  expressed as nominal wealth divided by the exergy price, would result in e.g. in $[USD]$ divided by $[USD/J]$, which reduces to $[J]$ as money cancels out:
\[\frac{[USD]}{[USD/J]}=[J]\]
Real economic growth would violate the first and second thermodynamic principles, increasing energy and exergy inside the economy without a corresponding input. Any other scalar material numeraire (e.g. a Gold or Big Mac numeraire) would violate the mass conservation law. Value is not gold nor Big Macs nor exergy. 
\par
Instead, exergy (i.e. the physical value unit) is \emph{proportional} to the numeraire (i.e. the economic value unit); the exergetic numeraire expresses real wealth as \emph{exergy value equivalent}. Value is not reducible to any other scalar physical quantity, just as e.g. temperature is not reducible to energy and light intensity is not reducible to power. If value is not reducible, then the value unit cannot be defined as a derived unit in terms of a \emph{formula}, but necessitates a distinctive base unit in \emph{language form}, similar to the other base units  of the metric (= SI) system (meter, kilogram, second, ampere, kelvin, mole and candela).

\section{4. The natural value unit and its interpretation}

The natural units of the universe are the Planck units. Planck proposed them in 1899 and called them \emph{natural} units as they are \textquotedblleft constant for all times and all civilisations, even for non-human ones\textquotedblright \cite{Planck1899}. They are defined as square roots of expressions containing the fundamental physical constants G (Newton's gravity constant), c (speed of light), $\hbar$ (Dirac constant),  k (Boltzmann constant) and $\kappa_C$ (Coulomb's force constant). Their most up to date CODATA values  can be found at $http://physics.nist.gov/cuu/index.html$. The Planck energy is defined as $\sqrt{\hbar c^5/G}$, equalling 1956.1 MJ, which makes it a unit of human scale. Planck units simplify theoretical physics. With the Planck unit c=1, EinsteinÕs famous mass equivalence of energy becomes numerically E = m, i.e. the Planck energy is numerically equal to the Planck mass, but mass and energy remain distinct concepts. With the unit k=1, the average energy per particle per degree of freedom is numerically equal to half the corresponding temperature. With k=1, information and statistical entropy differ numerically only by a factor ln2. 

\par 
The naming convention of the metric (= SI) system follows the practice to name units in honour of scientists who have made outstanding contributions in their respective research field. Leon Walras (1834 - 1910) has been the first to write a complete multi-equation model of general equilibrium and has proved that there exists exactly one numeraire at equilibrium. The proposed name and definition of the natural value unit are as follows:

\par
\vspace{10mm}
\begin{tabular}{|l|}
\hline
1 walras (Wal) is the value of $\sqrt{\hbar c^5/G}$ (=1956.1 MJ) energy available in an environment\\ 
 at physical and chemical equilibrium $<=>$ 1 Wal is the value of 1965.1 MJ exergy\\
\hline
\end{tabular}
\vspace{10mm}
\par 
In Romance languages and in English, the abbreviation \emph{Wal} reminds of value. 
\par 

In its magnitude, this natural value unit is of the size of many current economic transactions, as TABLE \ref{Table1} shows. This fact could in itself justify its adoption as an accounting unit. A standard transaction size is needed in case money reveals itself from empirical analysis to be a non-inear value scale. Such non-linearity of the money scale would be the logical consequence of quantity discounts and non-constant returns to scale. Thus a transaction of e.g. 1000 USD might a priori not have the same value as 10 transactions of 100 USD.

\begin{table}
\caption{\label{Table1}}
\begin{ruledtabular}
\begin{tabular}{|ll|}
\hline
The Planck energy =  $\sqrt{\hbar c^5/G}$ & = 1956.1 MJ \\
= 543 kWh electricity & = 1 year lamp burning at 62 Watt    \\
= 52 l (= 43 kg) diesel oil & = 59 l (= 44 kg) gasoline \\
= Slightly more than one car tank & = Slightly more than 1/3 barrel of oil \\
= 60 l (= 55 kg) olive oil & = 117 kg sugar \\
= 60 - 90 kg hard coal & = 150 - 200 l wood cuts  \\
\hline

 \end{tabular}
 \end{ruledtabular}
 \end{table}

\par

The Planck energy has a physiological interpretation as it is also equivalent to 1 year of uninterrupted energy consumption at a rate of 1280 kcal per day. This can be compared with the human metabolic rate, which depends on the four factors: Gender, age, weight and activity of the person. A consumption of 1280 kcal per day corresponds to the Resting Metabolic Rate (RMR) of a person of female gender, of the age category 18 - 30, whose weight is determined by resolving the equation $C = 14.7 W + 496$ for W, where C is the daily consumption (here: 1280 kcal) and W the weight in kg, cf.  Appendix on p. 134 of\cite{Smolin/Grosvenor2000}. We get a resulting weight of 53 kg. This category of person and its activity (i.e. at rest) perfectly corresponds to the Sleeping Beauty. The Planck energy can therefore be interpreted as her annual minimum physiological energy consumption. A value of one walras equals the value of this amount of energy, i.e. her annual minimal real cost of life. This physiological minimum is many times lower than the social minima of various countries or the average cost of living used as measurement unit in inflation indices.

\section{5. Hedonic Estimation of the PhPP of money}

The estimation of the physical purchasing power (PhPP) of money requires measuring the exergy as well as the market prices of all exergy-carrying commercially available goods. In principle, any good, no matter whether consumable or durable, carries exergy. Following the criteria of chapter 3, not all goods can however be numeraire. The numeraire must be a one-time consumable. This characteristic is strictly speaking not determined by the nature, but by the \emph{use} of a good. Only goods bought for their exergetic  \emph{use} are relevant. The exergetic use is the interactive property (or quality or characteristic) of interest of these goods. Besides that, different exergetic goods have different intrinsic properties which may all influence the good's price. These properties are not part of the numeraire. It is necessary to estimate the exergy price conditional on all other qualities or characteristics. The state of the art method to do this is hedonic regression. The term \emph{hedonic} has first appeared in\cite{Court1939}. Hedonic regressions are ever more used also in exact inflation measurement (axiometry), especially for estimating quality change\cite{Griliches1971}. The term axiometry has sometimes been used for designating exact inflation measurement\cite{Walsh1921}. Hedonic regressions can also be used for explaining price changes between specimens of a large variety of models as function of their main characteristics (characteristics model)\cite{Lancaster1966}\cite{Lancaster1971}. Hedonic regressions yield product-neutral parameters, in our case the product-neutral hedonic numeraire price and its inverse, the product-neutral physical purchasing power (PhPP) of money. The hedonic regression in this research uses transactions as observations. This guarantees the validity of the results for all observed transaction sizes, but the result might be non-linear and call for the definition of a \emph{standard transaction size} for comparing currencies. 
\par

Hereafter we summarize the results of the estimation of the Hedonic Walras Price (HWP) in Switzerland in 2003 and of its inverse, the Physical Purchasing Power (PhPP) of the Swiss Franc in 2003. The results are provisional because only one country and year have been studied yet, and only energy products have been included, neglecting food, feed and explosives. Both the hedonic Walras price (HWP) quoted in CHF and its inverse, the physical purchasing power PhPP of the CHF are quasi linear in transaction size. HWP at transaction size of 1 Wal was 102.36 CHF per Wal in 2003. In terms of the physiological interpretation, this means that the annual minimum real cost of life (i.e. of physiological energy consumption) of the Sleeping Beauty, if bought in one transaction, was 102.36 CHF in 2003. We propose to take the HWP of the Swiss franc in 2003 as conventional standard denominator transforming nominal 2003 wealth, expressed in CHF, to real wealth. The resulting dimension of real wealth is the Wal: 
\[\frac{[CHF]}{[CHF/Wal]}=[Wal]\]
\par 
Conversely, the PhPP of the Swiss franc at transaction size of 1 Wal was 0.01 Wal / CHF in 2003. The PhPP of the Swiss franc in 2003 is proposed as the conventional standard proportionality factor or exchange rate transforming nominal CHF 2003 wealth to real wealth, which again has the dimension Wal: 
\[[CHF]*\frac{[Wal]}{[CHF]}=[Wal]\]
\par 
With this convention, all nominal prices, including the price data set underlying this research can be transformed to real prices. This convention also allows keeping \emph{real} private, public and national accounts in terms of Wal. The real per capita GDP of Switzerland in 2003, estimated from nominal per capita income divided by HWP, amounts to 577 Wal per person per year, interpreted as 577 times the biological minimum real cost of life if acquired in transactions of size 1 Wal. Econophysics authors studying the physics of income distribution\cite{Dragulescu/Yakovenko2001}\cite{Mimkes2005} have found an analogy between the average annual per capita income of an economy and the temperature of a gas. The present contribution substantiates this analogy: If the income is taken as \emph{real} annual per capita income and expressed in Wal per person per year, it is directly proportional to energy (expressed in MJ) with proportionality factor 1956.1 MJ / Wal per person per year. The temperature of a gas, expressed in K, is known to be proportional to energy (expressed in J), too, with proportionality factor $k/2 = 0.6903252 E-23 J/K$ per particle per degree of freedom. Thus, the analogy between real income and temperature consists of the fact that they are both proportional to energy.

\section{6. Conclusion}

\par 
In this article we rejected the idea of value being an intrinsic or absolute quantity of goods. On the contrary, we defined the value of a good as being the result of an interaction between that good and a specific agent. Given that money, too, has a variable value called purchasing power, we treated money as a non-calibrated measurement instrument whose measurements need conversion to a fix unit. For conversion purposes, we rejected the usual consumer basket, as it is variable, imprecise and creates a circular measurement paradigm. Theoretical research shows that no ideal multilateral index exists that would give consistent results when consumer baskets are variable. We defined a natural value unit on the basis of a physical phenomenon and estimated the exchange rate of the Swiss Franc in 2003 in terms of this unit. Contrary to the Gold Standard, which was based on a man-made law, the exchange rate we found results only from the agents' behavior. We showed that our natural value unit has a physiological interpretation allowing to understand real income estimated by hedonic regression as a multiple of the annual cost of physiological life at minimum activity. This \emph{cost of life} has the advantage to be more precisely defined than the \emph{cost of living} calculated with traditional inflation indices. 

\par

\begin{appendix}
\appendix
\section{Appendix: Hedonic Regression}

\par 
The model underlying the estimation of the Hedonic Walras Price (HWP) and its inverse, the Physical Purchasing Power (PhPP) is the so-called covariance model, cf. p. 254 of\cite{Pindyck/Rubinfeld1988}. It regresses transaction prices P on the corresponding quantity of numeraire V, on other metric variables or characteristics $C_j$ and on one dummy variable $D_j$ for each energy product: 
$$ P= \alpha_0+\alpha_VV+\alpha_1C_1+\alpha_2C_2+...+\alpha_kC_k+\alpha_{k+1}D_1+...+\alpha_{k+p}D_p+  \varepsilon     $$  
The above model can be called price model, as the transaction price is the dependent variable. A dual form of this model exists, regressing the transaction numeraire V on the transaction price P, with $C_j$ and $D_j$ as other covariates:
$$ V= \beta_0+\beta_PP+\beta_1C_1+\beta_2C_2+...+\beta_kC_k+\beta_{k+1}D_1+...+\beta_{k+p}D_p+  \delta     $$  
The dual model can be called value-model, as the numeraire is the dependent variable. The primal and the dual models are related as follows:
$$ V= \frac{1}{\alpha_V} (P-\alpha_0 - \alpha_1C_1-\alpha_2C_2-...-\alpha_kC_k-\alpha_{k+1}D_1-...-\alpha_{k+p}D_p-  \varepsilon)     $$  
In the primal model, the marginal or hedonic walras price (HWP), which is our primal quantity of interest, is given by
$$p\textquoteright_V=\frac{\partial P}{\partial V}(C_1,C_2,...,C_k,D_1,...,D_P)  $$
Sometimes this marginal price is called marginal value\cite{Feenstra1995} or  implicit price, characteristics price,  hedonic price or \textquotedblleft shadow price\textquotedblright  \space\cite{Pollak1983}. The dual model naturally gives the physical purchasing power (PhPP) of the currency, which is our dual quantity of interest. It is given by
$$v\textquoteright_P=\frac{\partial V}{\partial P}(C_1,C_2,...,C_k,D_1,...,D_P)  $$
In the general non-linear case, PhPP is the inverse of the HWP only if V is monotonous in P and vice versa. In this case, one can estimate both models and retain whichever gives a better statistical fit. In linear models, $p\textquoteright_V$ is given by $\alpha_V$ and $v\textquoteright_P$ by $\beta_P$ with $\alpha_V=1/\beta_P$.
\par 
The estimation of HWP and PhPP of the Swiss Franc in 2003 has been carried out on the basis of 24'249 disaggregated energy price observations from the Swiss Federal Statistics Office\textquoteright s Consumer Price Index (CPI) and Producer Price Index (PPI) surveys for 2003. The meta-information of the CPI and PPI includes definitions of the consumer categories (for electricity, gas and district heating, e.g. for gas: 1000 kWh/year, 20'000 kWh/year, 50'000 kWh/year, 100'000 kWh/year, 500'000 kWh/year) and quantity bands (for heating oil) to which the observed prices apply. Transaction sizes have been constructed for all the metric variables (P, V and the mass M) on the basis of this meta-information and of experimental fuel tests\cite{SFOE1999} (for liquid fuels) or of publicly available information (for all the other energy products). Transaction sizes of grid bound energies (electricity, gas, district heating) are calculated for a standard contract duration (the same for all), which can be varied for improving statistical fit. Duration of 1 day has given the best fit. Hence it would in principle be possible to estimate a daily HWP and PhPP. For the cases where prices do not depend on transaction size (e.g. motor fuels and wood products), a single transaction size per product has been assumed. The provisionally best model is:
$$ln V= \gamma_0+\gamma_P lnP+\gamma_M lnM + \gamma_1D_1+...+\gamma_6D_6+\eta $$ 
In order to allow the natural log to be taken for all observations of all metric variables, all electricity transactions have been given a symbolic mass corresponding to the mass of one electron, independent of transaction size. The names of the co-variates, their coefficients\textquoteright \space  values, standard errors and respective t-ratios are given in TABLE \ref{Table2}. 

\begin{table}
\caption{\label{Table2}}
\begin{ruledtabular}
\begin{tabular}{|l|l|l|l|}
\hline
Co-variate & Mean value & Standard error & t-ratio \\
\hline
Constant (= electricity) & 2.76023 & 0.0993 & 27.8 \\
Ln of Price & 0.997242 & 0.0016 & 636  \\
Ln of Mass &0.143091 & 0.0019 & 76.5 \\
District Heating (Dummy) & -12.2567 & 0.1447 & -84.7 \\
Car Fuel (Dummy) & -10.1382 & 0.1345 & -75.4 \\
Extra Light Heating Oil (Dummy)  & -9.62431 & 0.1395 & -69.0  \\
Raw Wood (Dummy) & -9.52903 & 0.1430 & -66.6 \\
Gas ( Dummy) & -9.25687 & 0.1348 & -68.7 \\
Dry Wood (Dummy) & -8.38461 & 0.1435 & -58.4 \\
\hline

 \end{tabular}
 \end{ruledtabular}
 \end{table}

\par
TABLE \ref{Table2} orders the energy products by their quality, from district heating (lowest) to electricity (highest, having no mass). In the non-linear model, only the massless energy can be chosen as reference category. This provisionally best model is part of the class of models with adjusted $R^2 = 99.5$. Among these, this model is the one with the least number of predictors, i.e. with the coarsest grouping of dummy variables, so that only 6 from originally 15 of them remain. Hence, this model has 6 + 1 = 7 product categories. After having removed the logarithms, these results allow stating the provisional correspondence between transaction size, the hedonic numeraire price HWP and the physical purchasing power PhPP, all with respect to the reference category electricity, as shown in TABLE  \ref{Table3}. 

\begin{table}
\caption{\label{Table3}}
\begin{ruledtabular}
\begin{tabular}{|l|r|r|r|r|r|r|}
\hline
Transaction Size (Wal) & 0.01 & 0.1& 1 & 10 & 100 & 1'000 \\
\hline
HWP (CHF / Wal)  & 101.06 & 101.71 & 102.36 & 103.01 & 103.67 & 104.33  \\
PhPP (Wal / CHF) & 0.010 & 0.010 & 0.010 & 0.010 & 0.010 & 0.010 \\
\hline

 \end{tabular}
 \end{ruledtabular}
 \end{table}

\par



\end{appendix}
\begin{acknowledgments}

\end{acknowledgments}

\bibliography{defilla4}

\end{document}